\documentclass[12pt]{article}

\input epsf
\def\beq{\begin{eqnarray}}
\def\eeq{\end{eqnarray}}

\def\({\left(}
\def\){\right)}

\def\mpl{M_{\rm pl}}

\def\ea{\end{eqnarray}}
\def\ba{\begin{eqnarray}}

\def\beq{\begin{eqnarray}}
\def\eeq{\end{eqnarray}}

\def\({\left(}
\def\){\right)}
\def\mn{_{\mu \nu}}

\def\mpl{M_{\rm pl}}

\def\lsim{\mathrel{\rlap{\lower3pt\hbox{\hskip0pt$\sim$}}
     \raise1pt\hbox{$<$}}}         
\def\gsim{\mathrel{\rlap{\lower4pt\hbox{\hskip1pt$\sim$}}
     \raise1pt\hbox{$>$}}}         

\def\lsim{\mathrel{\rlap{\lower3pt\hbox{\hskip0pt$\sim$}}
     \raise1pt\hbox{$<$}}}         
\def\gsim{\mathrel{\rlap{\lower4pt\hbox{\hskip1pt$\sim$}}
     \raise1pt\hbox{$>$}}}         

\newcommand{\comment}[1]{}
\usepackage{amsmath}
\usepackage{amsfonts}
\usepackage{verbatim}
\usepackage{graphicx}
\usepackage{subfigure}
\usepackage{amssymb}
\usepackage[T1]{fontenc}

\renewcommand{\comment}[1]{}

\begin{document}

\centerline{\Large \bf Restricted Galileons}
\vskip 0.2cm
\centerline{\Large \bf }

\vskip 0.7cm
\centerline{\large Lasha  Berezhiani$^a$, Giga Chkareuli$^b$ and 
Gregory  Gabadadze$^b$}
\vskip 0.3cm

\centerline{\em $^a$Center for Particle Cosmology, Department of Physics and Astronomy,}
\centerline{\em University of Pennsylvania, Philadelphia, Pennsylvania 19104, USA}

\centerline{\em $^b$Center for Cosmology and Particle Physics,
Department of Physics,}
\centerline{\em New York University, New York,
NY, 10003}

\vskip 1.9cm

\begin{abstract}

We study Galileon theories that emerge in ghost-free 
massive gravity. In particular, we focus on  a sub-class of these theories 
where the Galileons  can be completely decoupled from the tensor Lagrangian. 
These Galileons differ from generic ones -- they have  
interrelated coefficients of the cubic and quartic terms, 
and most importantly, a non-standard coupling to external 
stress-tensors,  governed by the same coefficient.  
We show that this theory has no static stable spherically symmetric solutions that  
would interpolate from the Vainshtein region to flat space; these two regions cannot 
be smoothly matched for the sign of the coefficient for which fluctuations are stable.  
Instead, for this sign choice, a solution in the Vainshtein domain is matched 
onto a cosmological background. Small fluctuations above this solution are stable, 
and sub-luminal.  We discuss observational constraints on this theory, 
within the quantum effective Lagrangian approach, 
and argue that having a  graviton mass of the order of the present-day Hubble 
parameter, is  consistent with  the data.  Last but not least, we also present 
a general class of cosmological solutions in this theory, some of which  
exhibit the de-mixing phenomenon,  previously found 
for the self-accelerated solution.

\end{abstract}

\newpage

\section{Introduction and summary}

In the present work we will  be interested in those Galileons that
describe  helicity-0 modes of tensor fields:
In the context of the DGP model \cite {DGP} the cubic Galileon
was found in Ref. \cite {Rattazzi1}, and was shown  to describe properties 
of the helicity-0 mode. More general, quartic and quintic Galileons, 
were studied in \cite {Rattazzi2}.

Most relevant to the present work, is the fact that all the Galileons 
appear in the Lagrangian for the helicity-0 mode of 
a massive graviton \cite {dRG,dRGT} (see \cite {dRG0} 
for earlier evidence). Although the massive gravity Galileons are similar 
to the generic ones \cite {Rattazzi2}, there is a nuance that distinguishes 
them:  the  massive gravity Galileons  come  with specific 
coefficients,  and most importantly, with 
a novel coupling to the matter stress-tensor 
governed by the same coefficient \cite{dRG0,dRG,Wyman}. 

One motivation  for this  work  is  to 
study spherically symmetric solutions in this 
special class of massive gravity Galileons. 
In particular, in the present 
work we will focus on  the sub-class of  these 
theories for which the helicity-0 and helicity-2 
modes can be diagonalized \cite {dRG}. 
We refer to this sub-class as Restricted Galileons.

Spherically symmetric solutions 
in general Galileon theories are well understood \cite {Rattazzi1}:
generically, one gets a solution that transitions from the Vainshtein 
region to the asymptotically flat one; furthermore,
typically, stable fluctuations, both  in and outside of 
the Vainshtein region \cite {Arkady},  exhibit super-luminal propagation \cite {Rattazzi2}.
This however, does not necessarily lead to acausality -- 
attempts to create  closed time-like curves bring one beyond
the domain of validity of an effective field theory \cite {AndrewSuper}.
Pending a classical, or quantum, completion of these theories
at the strong scale, it makes sense to postpone the discussion of 
this issue. For a review of other interesting theoretical 
constructions with Galileons, see Refs. \cite {Mark}.

We will show that, there are significant 
differences in the case of restricted  Galileons (ReG). 
In particular, we will prove that in ReG 
no stable static spherically symmetric solutions exist 
that would interpolate between the Vainshtein region to 
an asymptotically flat domain. Instead, the Vainshtein region is 
naturally matched onto a solution that does not asymptote to 
flat space, but does to a cosmological background. 
The perturbations on this solution are stable, and sub-luminal! 

For  the exact static solution, 
we address the issue of the  bound on the graviton mass, as inferred from it. 
We  show that having the graviton mass as small as
the value of the Hubble parameter today is consistent
with the bounds derived from the table-top sub-millimeter measurements of 
gravity-competing forces. Our conclusions  differ from those 
in \cite {Nemanja}, and we discuss the origin of this difference.

The exact static solution, if applied to distribution of   
sources like the ones in our Universe (galaxies, clusters, etc), 
is consistent with the observations, due to the fact that 
the Vainshtein radius in such a universe gets pushed --
statistically speaking -- closer to the Hubble scale. There may 
however be isolated clusters in the universe for  
which beyond-the-Vainshtein-scale physics can be 
explored.

One could  ask what would happen to a dilute distribution of matter that  
initially has no Vainshtein  region, which is placed 
in an asymptotically flat space, and is  being  adiabatically
collapsed to form a smaller object with higher densities. 
Then,  the energy density and pressure in the nonlinear helicity-0 
field blow  up when the size of the distribution becomes of the order of 
the Vainshtein radius, preventing such an adiabatic collapse.  Thus, we'd 
expect the respective spherically symmetric collapsing solution, if it exists 
at all, to have time dependence that is fast as 
compared to the  Vainshtein scale.  In Appendix A we also present a class of  
more general time-dependent  solutions, however, non of them reach an 
asymptotically flat space for any constant time slice. 

Three important comments on the literature:  

(A) All the results of the present work are 
obtained in the decoupling limit. As to whether our results 
apply to the solutions of the full theory, discussed in Refs. 
\cite {Theo,Koyama1,Koyama2,Koyama3,AndreiMehrdad}, remains to be seen.

(B) Refs. \cite {Deser1,Deser2} have recently shown the existence of 
superluminal shocks for some parameter space of massive gravity. 
Such superluminal solutions, like in Galileon theories, appear
well-within the effective field theory domain. However,  
as briefly discussed above in the 4th paragraph of this section, 
and in detail in Ref. \cite {AndrewSuper}, the question 
whether the corresponding solutions can lead to acausal 
stable closed timelike curves, needs to be addressed including perturbations above these 
solutions, in the context of an effective field theory. In cases considered
so far, the fluctuations above such solutions bring one beyond the 
effective field theory \cite {AndrewSuper} -- one more argument that these theories need 
either calculational prescription, or classical/quantum completion below/at 
the strong scale\footnote{One example, where superluminality
was claimed to be removed, at least for spherically symmetric sources at the classical level, 
is a bi-galileon theory \cite {PadillaBi}. This requires a low energy remnant -- 
the second Galileon which is massless and operates at arbitrarily low scales. 
We note that certain bi-galileons naturally  emerge in generalization of ghost-free 
massive gravity \cite {dRGT} to a theory with a dilaton-like particle, quasidilaton 
\cite {Quasidilaton}. Although the issue of superluminality for the quasidilaton has not been 
studied yet,  predictions of this theory are not brought in conflict 
with observations by the presence of the low-energy remnant, the quasidilaton \cite{Quasidilaton}.}.

(C) Ref. \cite {Deser1} has also 
brought up 1970 lecture notes by Zumino \cite {Zumino}, 
where Wess and Zumino had proposed four models of a massive  spin-2 hadron, $f_{\mu\nu}$,
interacting with Einstein's gravity for $g_{\mu\nu}$.  Using the recently developed methods in 
\cite {dRG,dRGT,Rachel1,Rachel2,Mehrdad,RachelKurt}, we can tell now 
that  the two  (interrelated) ``f-g'' models  of Wess and Zumino  are  ghost-free as full  
non-linear theories and represent a subset of more general ghost-free bi-gravities \cite {Rachel2};  
if reduced to a  single massive graviton,  they represent a subset of  the  
ghost-free massive gravities \cite  {dRGT}, from which \cite {Rachel2} originates.  
The two other theories proposed in \cite {Zumino} have Boulware-Deser (BD)
ghosts \cite {BD}. None of the conclusions on the absence or presence of the 
BD ghost can  be  deduced from Ref. \cite {Zumino} itself, not surprisingly, 
as it precedes the BD work \cite {BD}. More details will be given elsewhere.

\section{The Theory}

In the decoupling limit of massive gravity, 
the Lagrangian describing  helicity-2 states, contained 
in $h_{\mu\nu}$, and helicity-0 state, denoted by $\pi$, with their 
coupling to matter stress-tensor $T_{\mu\nu}$, 
is given by \cite {dRG}:
\beq
\mathcal{L}= -\frac{1}{2} h^{\mu\nu}\mathcal{E}_{\mu\nu}^{\alpha\beta} h_{\alpha\beta} + 
h^{\mu\nu}\left (X^{(1)}_{\mu\nu}+\frac{\alpha}{\Lambda^{3}_{3}}
X^{(2)}_{\mu\nu} + \frac{\beta}{\Lambda^{6}_{3}}
X^{(3)}_{\mu\nu} \right ) + \frac{1}{ \mpl}h^{\mu\nu}T_{\mu\nu},
\label{lagr}
\eeq
where, the three conserved symmetric tensors  
$X^{(n)}_{\mu\nu}(\Pi)$ depend on the second derivatives of helicity-0 
field $\Pi_{\mu\nu}\equiv \partial_\mu \partial_\nu \pi$, and can be cast  
as follows \cite{dato},
\beq
X^{(1)}_{\mu\nu}&=&-\frac{1}{2}{\varepsilon_{\mu}}^{\alpha\rho\sigma}{{\varepsilon_\nu}^{\beta}}_{\rho\sigma}\Pi_{\alpha\beta}, \quad  \nonumber \\
X^{(2)}_{\mu\nu}&=&\frac{1}{2}{\varepsilon_{\mu}}^{\alpha\rho\gamma}{{\varepsilon_\nu}^{\beta\sigma}}_{\gamma}\Pi_{\alpha\beta}\Pi_{\rho\sigma},\nonumber \\
X^{(3)}_{\mu\nu}&=&\frac{1}{2}{\varepsilon_{\mu}}^{\alpha\rho\gamma}{{\varepsilon_\nu}^{\beta\sigma\tau}}\Pi_{\alpha\beta}\Pi_{\rho\sigma}\Pi_{\gamma\tau},\nonumber
\eeq
where  $\varepsilon_{\mu\nu\alpha\beta}$ is the usual Levi-Civita symbol.

Under the linear diffeomorphisms $\delta h_{\mu\nu} = \partial_{\mu}\zeta_{\nu} +
\partial_{\nu}\zeta_{\mu}$, the full nonlinear Lagrangian \eqref{lagr} is invariant up to a total derivative, while 
it is exactly invariant under the field space galilean transformations, $\delta \pi = v_\mu x^\mu$, with 
constant $v_\mu$ .

Due to the specific structure of this Lagrangian,  
the coefficients $\alpha$ and $\beta$ do  not get 
renormalized by quantum loops; hence, any choice of their 
values is technically natural \cite {dRGHP2}. 

For generic values of these coefficients
the helicity-2  and helicity-0 modes mix, as seen from \eqref{lagr};
this mixing cannot be undone at the full nonlinear level, unless
$\beta =0$ \cite{dRG}.  The choice $\beta =0$ corresponds to selecting 
a special relation, $\alpha_3 = -\alpha_4/4$, between the 
coefficients $\alpha_3$ and $\alpha_4$ of the cubic and quartic $K$ 
terms of the full massive theory \cite {dRGT}.

For $\beta=0$, which as already mentioned 
is a technically natural choice stable under loop corrections, 
there exists an invertible field redefinition 
\beq
h\mn\rightarrow \bar{h}\mn+\pi \eta\mn+\frac{\alpha}{\Lambda_3^3}\partial_\mu 
\pi \partial_\nu \pi,
\label{redef}
\eeq
that decouples the tensor  and scalar modes from each-other 
\cite{dRG}. The Lagrangian \eqref{lagr} then 
takes the form  ${\cal L} = {\cal L}_{\bar h}  +  {\cal L}_{\pi}  $, where 
$ {\cal L}_{\bar h} $ is the linearized Einstein-Hilbert  
action for  $\bar h$ minimally coupled to $T_{\mu\nu}$, 
while 
\beq
\mathcal{L}_\pi=\frac{3}{2}\pi\Box\pi+\frac{3}{2} \frac{\alpha}{\Lambda_3^3}
(\partial\pi)^2\Box \pi+\frac{1}{2}\frac{\alpha^2}{\Lambda_3^6}
(\partial\pi)^2([\Pi^2]-[\Pi]^2)\nonumber \\
+\frac{1}{\mpl}\pi T+\frac{\alpha}{\mpl\Lambda_3^3}
\partial_\mu \pi \partial_\nu \pi T^{\mu\nu}.
\label{diag}
\eeq
Here, square brackets $[.]$ denote the trace.  Note that the absolute value of the only free 
parameter $\alpha$ is immaterial,  as it can be 
absorbed into the scale $\Lambda_3$, however, the sign of $\alpha$ will be  crucial 
[rescaling $\Lambda_3$ would correspond to rescaling of the graviton mass;  
for comparisons with massive gravity, however, 
we will manifestly keep the parameter $\alpha$,  and unrescaled $\Lambda_3$ in what follows].

Thus, for $\beta=0$, the dynamics of the helicity-$0$ mode is similar 
to the cubic and quartic Galileons, but with some special  
coefficients in front of these terms; 
moreover, the last term in the second line in \eqref{diag} 
is a coupling to stress-tensor that is not usually considered in the Galileon theories, 
which however is present in massive gravity  \cite {dRG0,dRG,Wyman}, 
(it will plays a crucial role).
As mentioned in the previous section, we refer to \eqref{diag} 
as Restricted Galileons, and discuss this theory in the present work. 
The case with $\beta\neq 0$, which appears 
to be quantitatively  different, will be  discussed in Ref. \cite {SGII}.

As to the parameter $\alpha$, {\it a priori} it can have the either sign. 
However, the last term in \eqref {diag},  leads to classical renormalization of the 
kinetic term for $\pi$. For a negative sign of $\alpha$ this renormalization 
is negative,
and  for most of the reasonable sources it would overshoot the sign of the $\pi$ 
kinetic term.  Furthermore, we will show in Appendix A, 
that classical renormalization  of the $\pi$ kinetic term  due to 
the non-linear  terms in \eqref{diag}, although  positive,  is in fact 
subdominant. Therefore $\alpha<0$ case 
is not physical: when a localized source (with density greater than $\Lambda_3^3\mpl$) 
is approached from far away,  at some point the $\pi$ field  will have vanishing  kinetic term,  
signaling an  infinitely strongly coupled regime; at yet shorted distances the kinetic term 
would flip its sign (if one could pass into this region), to convert $\pi$ into 
a ghost.  For this reason, our physical choice will be  $\alpha>0$.

\subsection{Spherically symmetric solutions}

In this work we study a gravitational field configuration created by 
a source of finite size $R$,  and a uniform density $\rho$. The pressure 
of the source will also have some important consequences,  
and will be discussed in Appendix B.

Without the loss of generality, the static spherically symmetric configuration 
can be parametrized by  the following ansatz for the metric perturbations
\beq
h_{00}=a(r), \qquad h_{ij}=f(r)\delta_{ij},
\label{anz}
\eeq
while for the helicity-0 we begin  by assuming the radial  ansatz $ \pi = \pi(r)$.
Then, the equations of motion for graviton reduces to two ordinary differential 
equations (this is after integrating them once and requiring the solution to 
vanish at the origin)
\beq
\label{1}
rf^\prime=-\frac{2M}{\mpl r}+\Lambda_3^3 r^2\lambda(1-\alpha\lambda) ,
~~~~ra^\prime=-\frac{2M}{\mpl r}-\Lambda_3^3 r^2\lambda\,,
\label{2}
\eeq
where above, and in what follows we use the notations for $\lambda$ and the Vainshtein radius $r_*$:
\beq
\lambda\equiv \frac{\pi'}{\Lambda^3_3 r}, \qquad r_*\equiv 
\( \frac{M}{\mpl^2 m^2•} \)^{1/3},
\eeq
and the prime denotes differentiation w.r.t.  the radial coordinate $r$.

The closed form of the equations of motion for the helicity-0  mode can be obtained by 
integrating out $a$ and $f$, using \eqref{1}. The net 
result reads as follows \cite {GigaDato}
\beq
3\lambda-6\alpha \lambda^2+2 \alpha^2\lambda^3 
= \left\{ \begin{array}{ll}
2\left(\frac{r_*}{r•}\right)^3
& \mbox{r>R} \\ 2\left(\frac{r_*}{R}\right)^3 
& \mbox{r<R} \\ \end{array} 
\right. .
\label{3}
\eeq
We could have also derived this latter  equation  by simply integrating 
the Galileon equations of \eqref{diag}, with the same initial conditions.

It is now straightforward to show that for $\alpha >0$ equation \eqref{3} 
has no solution that could interpolate from the Vainshtein region 
to an asymptotically flat one, i.e., 
to the region where $\lambda \to 0$. 
To see this we look at the cubic polynomial  in $\lambda$ at the 
lhs of \eqref{3}. The necessary and sufficient 
condition for the solution of \eqref {3} to interpolate between the Vainshtein region
and the asymptotically flat one,  is that for the 
polynomial on the lhs of \eqref{3} to have  only one zero, at $\lambda=0$ \cite {Rattazzi2}. 
However, this condition is not satisfied if $\alpha>0$:  
there are three real zeros  in this case, 
$\lambda_1=0, \lambda_{2,3} =(3 \pm \sqrt{3})/2\alpha$.  This  means 
that the solution that starts off in the Vainshtein region,  where $\lambda\gg 1$,  
matches onto  a solution that at $r\to \infty$ tends to 
$\lambda_2= {3 +\sqrt{3}\over 2\alpha}$.
The latter implies that asymptotically $\pi \to \Lambda_3^3 r^2 \lambda_2$.

Then, the solution to \eqref{1}-\eqref{3} for $\alpha>0$, 
in the leading approximation, is:

\textit{Outside the Vainshtein radius}
\beq
\lambda\simeq \frac{3+\sqrt{3}}{2\alpha}, \quad a\simeq -\frac{\Lambda_3^3r^2}
{2}\lambda, \quad f\simeq \frac{\Lambda_3^3 r^2}{2}\lambda (1-\alpha \lambda).
\label{outside}
\eeq

\textit{Inside the Vainshtein radius}
\beq
\lambda\simeq \frac{r_*}{\alpha^{2/3}r}+\frac{1}{\alpha}+\frac{r}{2\alpha^{4/3}r_*}, \quad a\simeq f\simeq - \frac {2M}{\mpl r}+
{\cal O}(r^2\lambda).
\label{largelambda}
\eeq
This solution describes the static spherically symmetric source of mass $M$ on a cosmological 
background, with the equation of state $p/\rho=-2/3+1/\sqrt{3}\approx -0.1$.

Next, we would like to proceed with the analysis of fluctuations around this solution. 
As it is clear from the above,  we gave up the asymptotic flatness. 
Therefore, it is natural to begin by studying  the region far away from the source.
Due to the decoupling between $\bar h$ and $\pi$ we can analyze their fluctuations separately.
Those of $\bar h$ are linear GR perturbations on a cosmological background 
with the above-mentioned equation of state, taken in the high momentum 
approximation (momenta being larger than the Hubble parameter, which in this case 
is $\sim m$). These fluctuations are stable, and there is no novelty in this part.  

Let us then look at the fluctuations of $\pi$, which we decompose as 
$\pi = \pi_{cl}(r)  + \sigma(x,t)$, where  $\pi_{cl}(r)$ denotes the classical background just
found in \eqref{outside}.  Appendix A gives an expression for the Lagrangian of 
fluctuations about a general background $\Phi$; substituting $\Phi = \pi_{cl}$ 
into \eqref{pert}, we obtain the quadratic Lagrangian 
for fluctuations around \eqref{outside}:
\beq
\mathcal{L}=3\left[ 5+3\sqrt{3} \right] (\partial_t \sigma)^2-
\frac{3}{2}\left[ 1+\sqrt{3} \right](\partial_j \sigma)^2.
\eeq
Hence, the cosmological background is stable against all linear perturbations. 
Moreover, the fluctuations propagate with  sub-luminal velocities. 

As a next step we study fluctuations in the presence of the
source of mass $M$. Outside the Vainshtein radius the just-studied 
cosmological background 
dominates, and hence the fluctuations are stable and sub-luminal. 
However, inside the Vainshtein radius the source defines the background 
and its effects are dominant.
The fluctuations of $\bar h$ coincide, to a good approximation, 
with those of linearized GR on a background of a static spherically 
symmetric source.  The fluctuations of $\pi$, however, need  an extra caution.
Using again \eqref{pert},  we find that in the leading order in $r/r_*$ 
(which is a good approximation in the Vainshtein regime where $(r/r_*) \ll 1$), the small 
excitations  around \eqref{largelambda} 
are described by the following Lagrangian: 
\beq
\mathcal{L}=3\left[ \alpha^{2/3} \left(\frac{r_*}{r}\right)^2+2\alpha^{1/3}\frac{r_*}{r} \right] 
(\partial_t \sigma)^2-3 \alpha^{2/3}\left(\frac{r_*}{r}\right)^2(\partial_r \sigma)^2-\frac{3}{2}
(\partial_\Omega \sigma)^2.
\eeq
Remarkably, radial fluctuations propagate with a slightly 
sub-luminal speed --the deviation from the luminality being suppressed by $r/r_*$. 
The angular speed of sound, on the other hand, is strongly suppressed.
This suppression,  for a generic quartic Galileons,  was already emphasized  
in Ref. \cite {Rattazzi2}.

It is interesting to ask what happens with a dilute distribution of matter 
with some size $R$,  and density below $\Lambda_3^3 \mpl$,   placed 
on an asymptotically flat space. Such a  source has no Vainshtein region, since $r_* <R$, and there is 
a static solution with decaying $\pi$ field at infinity. However, if the source is adiabatically
collapsed to form a smaller object with density above $\Lambda_3^3 \mpl$ (i.e., with 
$r_*>R$), then the energy density and pressure in the nonlinear $\pi$ field blows 
up when $r_* \sim R$, preventing such an adiabatic collapse.

\section{Observational prospects ?}

In this section we will consider a hypothetical universe described by the 
massive gravity with $\beta =0$. This is a technically natural choice, as it's not 
ruined, to a high degree of accuracy, by loop corrections \cite {dRGHP2}.
We take the solution \eqref{outside} as our background in an empty space, 
anticipating that localized sources will match onto this solution, and 
not to a solution with flat asymptotics.

\subsection{Nonperturbative effects in Solar System}

Since the background \eqref{outside} has an unusual equation of state,  
we will require  this fluid to be a subdominant component in the universe
\footnote{Although, other more interesting options may also be possible 
when the density of this fluid adds in a meaningful way to the matter 
density, while its 
small negative pressure, adds a bit to the negative pressure due to 
dark energy.}. This can be arranged by choosing the parameter $\alpha$ to be 
of order ten or so (for a fixed graviton mass $m\sim H_0$), 
which again, is a technically natural choice \cite {dRGHP2}.
Then, to obtain realistic universe we would have to introduce dark energy,  and 
matter density,  roughly as we do in the $\Lambda$CDM model. Therefore, from 
the point of view of getting dark energy from modified gravity, 
the $\beta =0$ theory is not interesting, however, our goal here 
is to understand if such a theory, in principle, could be consistent 
with observational data.

To this end we would like to discuss  observational constraints due to the small 
corrections to the GR results within the Vainshtein radius 
\cite{Arkady, Lue,Gruzinov}. From the solution \eqref{largelambda} we easily 
deduce that the fractional increase of the Newtonian potential inside
the Vainshtein radius is
\beq
q\equiv {\Delta \phi \over \phi} \simeq {m^2 r_* r /\alpha^{2/3} \over r_g/r},
\label{q}
\eeq
where by $\phi$ we denote  the Newtonian potential.
This deviation from the standard potential will give rise to 
an additional perihelion advancement of orbiting bodies 
(an additional to the GR effects).  For careful account 
of these effects, within a relativistic theory, see \cite {Lue,Iorio}.

Then, the question is whether such deviations can be measured.
One of the most precisely measured trajectory is that of the Moon
orbiting the Earth \cite {LLR}. 
For the Earth-Moon system, and for $m\simeq H_0$, the fraction \eqref {q}  
is of order $q^{Earth-Moon} =10^{-16}/\alpha^{2/3}$;  this is four 
orders of magnitude below  the same fraction in the DGP model \cite {Lue,Gruzinov}
(see, also \cite {GIglesias}); 
as such its influence on the perihelion advancement of the Lunar orbit is 
completely negligible and is unlikely to be measurable by the near 
future Lunar Laser Ranging  Experiments \cite {LLR}.

Next, we discuss whether fluctuations of the $\pi$ field, 
above its background,  can give rise to additional forces competing with gravity.  
There is an important point, in these considerations, as compared to conventional Galileons.
The kinetic term for the fluctuations $\sigma = \pi -\pi_{cl}$ 
gets additional classical renormalization
\beq
-\left (\eta_{\mu\nu} + Z^V_{\mu\nu} + {\alpha T_{\mu\nu} \over \mpl\Lambda_3^3}  \right )
\partial^\mu\sigma\partial^\nu\sigma+ \frac{1}{\mpl}\sigma T,
\label{sigmaT}
\eeq
where the factor $Z^V$  is due to the cubic and quartic 
Galileon  terms evaluated on the solution; 
we have also manifestly shown couplings of the fluctuations to the 
stress-tensor. Note that the kinetic term 
in \eqref{sigmaT} gets enhanced by both 
a factor due to the classical  background, $Z^V$, and by a source-dependent 
term $\alpha T$ (both of these are huge factors  for realistic sources, but 
as shown  in Appendix  A, the enhancement due to the $\alpha T$ 
term dominates within a source). As a result, the static force due to exchange of $\sigma$, 
between any two objects, say two small metallic balls, is suppressed by a tremendous factor  
determined by the ratio of density  and pressure in the balls over the density
$\Lambda_3^3 \mpl$; the latter happens to coincide with the critical density, $\rho_c \simeq 10^{-29}g/cm^3$, 
if $m\sim  H_0\simeq 10^{-33}~eV$. This fraction is  enormous, $(\rho_{metall}/\Lambda_3^3 \mpl)\gsim 10^{30}$.
As we will discuss  in the next section, this suppression 
makes the $\sigma T$ coupling unobservable. Therefore,  the exchange due to a single $\sigma$ field  
cannot be gravity-competing force. In the next section we turn to  quantum effects, 
and the question whether they could ruin the above conclusion.

\subsection{Comments on quantum effects}

In this section we would like to discuss  quantum loop effect.
Before we do so, one should  point out that we're dealing with 
a theory that is non-renormalizable  by power-counting, and at least in this sense, 
requires a completion at the scale $\Lambda_3$. However, judging from the properties of the 
theory below $\Lambda_3$ this completion, unlike in massive non-Abelian vector fields, 
is not expected to be a conventional one  (see \cite {Adams, dRGHP2}, and references therein). 
On the other hand, discussions  in the literature, as well as below,  are based on an 
assumption of a certain conventional generic completion. This is an assumption that may not 
be a right one, and  therefore, all such  discussions are on a shaky  ground.

Putting the above  concern aside,  we would  like to analyze  the  
viability of the bound imposed on graviton mass in Ref. \cite {Nemanja},  based on 
the conventional effective field theory considerations of the decoupling limit 
Lagrangian \cite {RattazziNicolis}.  
To do so we first briefly recall what do the quantum corrections lead to in the decoupling limit of 
massive gravity \cite {dRGHP2}: the unambiguous result, independent of the uncertainties of the 
previous paragraph, is that  the terms presented in \eqref{lagr} do not get renormalized!
However, in a conventional quantum effective field theory approach new terms may get 
induced in the 1PI action.  These new terms include  ambiguous power-divergent terms, 
as well as log divergent pieces that carry information 
about the forward scattering  of the quanta, and  have to be 
included in the 1PI action.  What was argued for the  cubic \cite {RattazziNicolis}, 
as well as generic Galileons \cite {Rattazzi2},  is that these terms, in the conventional 
approach,  and once calculated on a classical background, end up being suppressed
by a scale of the classical background itself.

The classical renormalization \eqref{sigmaT} increases the effective energy/momentum 
scale of strong interactions of the fluctuations in the 
1PI action \cite {RattazziNicolis}. The naive analysis shows that the  
loop-induced terms are suppressed by the effective scale 
schematically written as follows: 
\beq
\Lambda_{eff} \simeq  (Z^V + {\alpha T \over \mpl\Lambda_3^3 })^{1/3} 
\Lambda_3 \equiv Z^{1/3}_{tot} \Lambda_3.
\label{Lambdaeff}
\eeq
This is valid  when the enhancements of the kinetic and gradient terms are of the same order; however, 
\eqref{Lambdaeff} fails to describe the strong coupling scale in case of the hierarchy among them. For the 
illustration, consider the following toy model, that captures essential features of our theory 
on a static  background:
\beq
\mathcal{L}=a^2(\partial_t \pi)^2-b^2(\partial_j \pi)^2+\frac{b_*}{\Lambda_3^3}(\partial\pi)^2\Box\pi,
\label{toyab}
\eeq
where $a,b$ and $b_*$ are assumed to be constants for simplicity. The kinetic term for the 
field $\pi$ needs to be canonically re-normalized,  $\pi\rightarrow \pi/a$. This results in the 
naive $\Lambda_{eff}\sim a b_*^{-1/3} \Lambda_3$ estimate for the effective cut-off, in the spirit 
of \eqref{Lambdaeff}. However, by calculating the $2\rightarrow 2$ scattering amplitude one 
can see that the strong coupling scale  differs in different channels; this is because the  
propagator has the denominator, 
$\omega^2 - (b/a)^2 {\vec k}^2 $.  In fact, for the forward 
scattering and $b\ll a$, the $u$-channel is the strongest. 
It gives rise to the $\Lambda_{eff}\sim (b/b_*)^{1/3}a^{2/3}\Lambda_3$ cut-off, 
while the $s$-channel amplitude becomes 
strong only at $ab_*^{-1/3}\Lambda_3$ scale; the 
latter being considerably larger than the former. 

Higher order tree-level diagrams give higher effective scale: the vertices have more powers of 
inverse $a$ than the propagators can compensate for due to  the $(b/a)^2$ factor 
in the diagrams that aren't automatically zero\footnote{The power divergent terms in 
the loops, that could induce lower cutoff,  
are removed by the counterterms as per the general philosophy adopted in this approach;
the log divergent terms however, are related to the tree-level diagrams 
by unitarity and analyticity. Hence, it suffices to get the lowest cutoff of the 
tree-level diagrams as we did above.}.
All this remains to be the case even after  the quartic Galileon is  included in \eqref{toyab}. 
Hence, the effective lowest strong coupling scale remains to be:
\beq
\Lambda_{eff} = a^{2/3}  \left ( b\over b_* \right )^{1/3} \Lambda_3.
\label{effab} 
\eeq
At this scale, interactions of the fluctuations would become 
non-perturbative. Let us now discuss how this squares with the fact that 
gravity-competing forces have  been excluded down to the scales of order 100 
microns or so \cite {Adelberger}. First, recall that in the DGP model, 
$\Lambda^{-1}_{eff}$ is of order a centimeter \cite {RattazziNicolis},  
the scale much bigger then 100 microns. However, as pointed out in \cite {RattazziNicolis}
this is not likely to be an issue,  since these fluctuations are very weakly coupled 
to matter sources, and thus, their effects are unlikely to be visible in the submillimeter 
measurements, even though they themselves self-interact strongly at scales below 1cm. 

For the model considered  in the present work, 
it was argued in Ref. \cite {Nemanja} that the $\Lambda_{eff}^{-1}$, is much 
bigger that a millimeter, if the graviton Compton wavelength is taken to 
be of the Hubble size. Does this mean that  such a small graviton  mass 
is  ruled out by laboratory measurements, as claimed in \cite {Nemanja}?  

To address this question in detail,  consider a device 
measuring putative gravity-competing forces --  
a torque pendulum of the Adelberger's group experiment \cite {Adelberger}. 
Then there are two important points: 

(1) Within  the matter that the pendulum 
plates are made of  the effective strong coupling scale differs 
significantly from its value outside of the plates. 
Using \eqref {effab}, and the fact that in this case
$a\sim 10^{15}$, $b\sim b_*$, we find that in the plates 
$\Lambda^{-1}_{eff}\simeq 10^{-10}\, 1000~km \simeq 0.1 ~mm$.
This is smaller than the thickness of the  upper and lower pendulum plates, that  
respectively are  about 1.8 and 7.8 mm \cite{Adelberger}
\footnote{This scale is greater  than the width of the 
plates, 0.3 microns,  used in the experiment of Ref. \cite {Kapitulnik} that 
exclude forces 14,000 time stronger than gravity at 10 microns. However, 
the coupling to matter that we're discussing  is $10^{-18}$ of the coupling of 
gravity at 0.1mm (see below), and it would be unreasonable to expect for 
it to become 14,000 times stronger than gravity at the  micron scale.}.

(2) Thus, in the torque pendulum plates, the fluctuations are weakly coupled 
to themselves,  and they can  be well-described by the classical Lagrangian  
down to 0.1 millimeters. On the other hand,  as shown in the previous section, 
this classical Lagrangian gives rise to a tremendous suppression of 
the coupling $\sigma T/\mpl$, that determines how strong/weak 
the coupling of the fluctuations to the plates could be.   
For the metallic plates this suppresses  the $\sigma T/\mpl$   
vertex  by an additional  factor of $a^{-1} \sim 10^{-15}$!

Therefore, a single $\sigma$ fluctuation cannot efficiently be 
emitted by the plates (or any realistic source bigger that 0.1 millimeter, 
for that matter).  Even though the fluctuations  in the vacuum between the 
plates  become strongly interacting with themselves, and could form some 
bound states of $\sigma$, what's important is that this exotic strongly 
self-coupled sector cannot couple  efficiently to the measuring device -- the 
latter coupling  due to the exchange of $\sigma$ is $a^{-2} (b/a)^{-2} \sim 10^{-18}$ 
of the strength of gravity. Therefore,  the exchange due to $\sigma$  
cannot be gravity-competing force -- in spite of the fact that the $\sigma$ 
field self-interacts strongly outside of the plates. 
This suppression was not taken into account in Ref. \cite {Nemanja}, 
making the bound imposed on the graviton mass by that work unwarranted.
The valid phenomenological bounds on the graviton mass are given  in Ref. 
\cite {Fred}.

Note that  for $\beta \neq 0$, which was also considered in \cite {Nemanja},
the effect of the suppression of coupling to matter,  due to the $\partial^\mu 
\sigma \partial^\nu \sigma T_{\mu\nu}$ term, remains valid. However, there are technical 
and conceptual differences for $\beta\neq 0$ \cite {SGII}, that obfuscate the 
results of  Ref. \cite {Nemanja} for that case;  this and related issues 
will be presented in detail in \cite {SGII}.

In conclusion of this section, and to reiterate, we  note that many 
statements about the quantum effective theory are based  on 
assumptions about what the UV completion might be.  It is   
clear from various standpoints (see, e.g., \cite {dRGHP2} and references 
therein), that theories of massive gravity need a completion, 
or alternatively,  a dual formulation (perhaps along the lines 
of \cite {GHP,Padilla}), at and above the $\Lambda_3$ scale.
However, judging from the properties of these theories (such as 
e.g., nonrenormalizability of some of the couplings \cite {dRGHP2}) 
the  putative completion  should  not be expected to be of a 
conventional and generic type, as preassumed in  all the above discussions.  
Therefore,  conclusions deduced from such  considerations \cite {Nemanja}, 
even if they were plausible, would not be set in stones.

\vspace{1in}

{\bf Acknowledgments}

\vspace{0.1in}

We would like to thank Claudia de Rham, Rampei Kimura, Mehrdad Mirbabayi, Rachel A. Rosen, Andrew Tolley and Arkady 
Vainshtein for discussions and helpful comments on the manuscript.  LB is supported by funds provided by the University 
of Pennsylvania, GG is supported by NSF grant PHY-0758032 and NASA grant
NNX12AF86G S06.

\vspace{0.5in}

\renewcommand{\theequation}{A-\Roman{equation}}
\setcounter{equation}{0} 

\section*{Appendix A}

In order to study the stability, we split the scalar degree of freedom into the 
background $\Phi$ and the fluctuation $\sigma$ as follows
\beq
\pi=\Phi+\sigma.
\eeq
As a result, to the second order in perturbations, the Lagrangian \eqref{diag}  becomes\footnote{Here we give only the scalar part of the Lagrangian, since the tensor mode is completely decoupled and propagates according to the linearized Einstein-Hilbert action.}
\beq
&&\mathcal{L}_{\sigma}=\left\{\left[ -\frac{3}{2}+3\frac{\alpha}{\Lambda_3^3}\Box\Phi+\frac{3}{2}\frac{\alpha^2}{\Lambda_3^6}\left( (\partial_\alpha \partial_\beta\Phi)^2-(\Box \Phi)^2 \right) \right]\eta_{\mu\nu}\right.\nonumber\\
&&\left.+\left[-3\frac{\alpha}{\Lambda_3^3}\partial_\mu\partial_\nu\Phi+3\frac{\alpha^2}{\Lambda_3^6}\left(-\partial_\alpha\partial_\mu\Phi \partial_\alpha\partial_\nu\Phi+\Box\Phi\partial_\mu\partial_\nu\Phi \right) \right]\right\}\nonumber\\
&&\times\partial_\mu\sigma\partial_\nu\sigma+\frac{1}{\mpl}\sigma T+ \frac{\alpha}{\mpl\Lambda_3^3}\partial_\mu \sigma \partial_\nu \sigma T^{\mu\nu}.
\label{pert}
\eeq
For the static clump of dust of constant density, that is for $T_{\mu\nu}=\rho \delta_{\mu}^0\delta_{\nu}^0\theta(R-r)$, we obtain the following kinetic term (to the leading order) inside the source
\beq
\left[\alpha\frac{\rho}{\mpl\Lambda_3^3}+9\left(\frac{\alpha^{1/3} r_*}{R}\right)^2\right](\partial_t\sigma)^2.
\label{ghost}
\eeq
Here, the second term in brackets comes from the quartic Galileon term after we have substituted the classical solution within Vainshtein region \eqref{largelambda}.
After the replacement $\rho\rightarrow M/R^3$, \eqref{ghost} leads to the conclusion that in $\alpha<0$ parameter space the scalar perturbations have ghost-like kinetic term within the source
of radius $R\ll \alpha^{1/3}r_*$. The latter condition can be translated on the language of the source density as
\beq
\alpha\rho>\mpl\Lambda_3^3.
\eeq
For phenomenologically interesting value of $\Lambda_3^{-1}\sim 1000~\text{km}$ the above bound becomes $\alpha\rho>10^{-29}~\text{g}/\text{cm}^3$. 

As we've shown, for $\alpha >0$, the asymptotically flat solution cannot be continued
into the Vainshtein region. However, there are cosmological solutions that are free of the
above difficulty.

In Section 2 we have already presented the static solution which asymptotes to the cosmological background, rather than the Minkowski space. Here we consider the time-dependent ansatz:
\beq
\pi=\frac{c}{2}\Lambda_3^3t^2+\pi_0(r),
\label{pianz}
\eeq
The introduction of non-zero $c$ modifies \eqref{3} in the following way
\beq
&&3(1+2\alpha c)\lambda-6(\alpha+\alpha^2 c )\lambda^2+2 \alpha^2\lambda^3 \nonumber
\\&&= \left\{ \begin{array}{ll}
2\left(\frac{r_*}{r•}\right)^3 (1+2\alpha c)+c,
& \mbox{r>R} \\ 2\left(\frac{r_*}{R}\right)^3 (1+2\alpha c)+c,
& \mbox{r<R} \\ \end{array} 
\right .
\label{3a}
\eeq
The above-mentioned static cosmological background corresponds to $c=0$ in \eqref{pianz}. However, there is a plethora of available backgrounds, which correspond to different values of $c$ and the only discriminating principle among various backgrounds must be their stability. In this letter, we do not pursue the detailed analysis of all the possible solutions. Merely, we would like to present one more solution which in our opinion is quite special.

From \eqref{3a}, it follows that there is a peculiar value of $c=-1/(2\alpha)$, which makes the equation independent of the mass of the source. As a result, eq.\eqref{3a} has three solutions. One of them being $\lambda=1/(2\alpha)$, which corresponds to an unstable self-induced de Sitter space found in \cite{dato} (it propagates a ghost) and we discard it. The other two are quite interesting
\beq
\lambda=\frac{1\pm \sqrt{3}}{2\alpha},
\eeq
in particular the pion cloud has zero pressure and positive energy. The easiest way to see this, is to look at the metric itself
\beq
a=f=\frac{2M}{\mpl r}-\frac{\Lambda_3^3}{4\alpha} r^2.
\eeq
The effective pressure takes the following form
\beq
p=\frac{\mpl}{3}G_i^i=-\frac{\mpl}{6}{\varepsilon_{i}}^{\alpha\rho\gamma}{{\varepsilon^i}^{\beta\sigma}}_{\gamma}\partial_\alpha\partial_\beta h_{\rho\sigma}=\frac{\mpl}{3}\Delta(f-a)=0.
\eeq
Moreover, the energy density of the pion fluid, filling the space, is given by
\beq
\rho=\mpl G_{00}=-\mpl \Delta f=\frac{3\mpl\Lambda_3^3}{2\alpha}>0, \qquad \text{for} \qquad \alpha>0.
\eeq
On this background, the quadratic Lagrangian for perturbations is
\beq
\mathcal{L}=6(\partial_t\sigma)^2-\frac{3}{2}(\partial_j\sigma)^2,
\eeq
everywhere in the space. As it is easy to see, the fluctuations are stable and propagate with the half speed of light. 

We would like to emphasize that the $\pi$ profile is independent of the source
\beq
\pi=\frac{\Lambda_3^3}{4\alpha}\left[-t^2+\left(1+\sqrt{3}\right)r^2\right].
\eeq
This means, that the presence of the static, spherically symmetric and pressure-less source does not excite the longitudinal mode of the massive graviton.

\renewcommand{\theequation}{B-\Roman{equation}}
\setcounter{equation}{0} 

\section*{Appendix B}

Here, we would like to discuss the effect of pressure ($p\ll \rho$) on the background and its stability. The reason is simple, most of the sources in the universe possess small, yet nonzero pressure.

For simplicity, we treat only $c=0$ case here, since the generalization is straightforward. The equation of motion for the longitudinal mode \eqref{3} takes the following form
\beq
&&3\lambda-6\alpha\lambda^2+2 \alpha^2\lambda^3 \nonumber
\\&&= \left\{ \begin{array}{ll}
2\left(\frac{r_0}{r•}\right)^3
& \mbox{Outside the source} \\ 2\left(\frac{r_*}{R}\right)^3\left[ 1+\frac{p}{\rho}\left(\alpha\lambda-3\right) \right]
& \mbox{Inside the source} \\ \end{array} 
\right .,
\eeq
where $r_0$ is an integration constant, which is determined by the matching condition at the surface of the 
source. As expected, the only change in the field configuration is the effective mass.

Let us recall that we have chosen the sign of $\alpha$ to be positive to avoid ghosts inside the source. 
The danger was coming from the last term of \eqref{pert}. But, if we avoid the ghost by 
appropriate choice of $\alpha$ then for nonzero pressure that term leads to the tachyonic 
contribution to the gradient energy. However, after taking a closer look, it can be shown 
that the healthy contribution to the gradient energy, coming from the quartic Galileon, 
overwhelms the unhealthy one; rendering the background stable. A simple estimate shows that 
this remains to be the case as long as, $\rho \gg p$.  For $p\sim \rho$,  on the other  hand,
the coefficient of the gradient term of the perturbations is proportional to  $(\rho - 2p)/p$,
and for $p> \rho /2$ flips the sign to yield gradient instabilities. Such instabilities
should manifest themselves in production of the $\pi$ quanta (albeit the coupling of these quanta to 
the matter is very weak).  Whether this may be dangerous, or interesting, for high density 
astrophysical objects, such as neutron stars, should be investigated including the fact
that they're rotating and are neither exactly spherically symmetric, nor homogeneous; that is, they lack properties used in the derivation of above-mentioned results.

\end {document}